# Performance Bounds on Sparse Representations Using Redundant Frames

Mehmet Akçakaya and Vahid Tarokh

*Abstract*— We consider approximations of signals by the elements of a frame in a complex vector space of dimension $N$ and formulate both the noiseless and the noisy sparse representation problems. The noiseless representation problem is to find sparse representations of a signal r given that such representations exist. In this case, we explicitly construct a frame, referred to as the Vandermonde frame, for which the noiseless sparse representation problem can be solved uniquely using $O(N^2)$ operations, as long as the number of non-zero coefficients in the sparse representation of r is $\epsilon N$ for some $0 \leq \epsilon \leq 0.5$, thus improving on a result of Candes and Tao [3]. We also show that $\epsilon \leq 0.5$ cannot be relaxed without violating uniqueness.

The noisy sparse representation problem is to find sparse representations of a signal r satisfying a distortion criterion. In this case, we establish a lower bound on the trade-off between the sparsity of the representation, the underlying distortion and the redundancy of any given frame.

*Index Terms*— Frames, sparse representations, redundancy, sparsity, distortion

## I. INTRODUCTION

LET r be a complex $N$ dimensional signal and $\mathcal{B}$ be a basis for $\mathbb{C}^N$. Then it is well-known that r has a unique expansion in terms of the elements of this basis. In particular, if $\mathcal{B}$ is the Fourier basis, then fast algorithms for computing the expansion coefficients of r are very well-known.

Consider now a set $\mathcal{F}$ of $M \geq N$ non-zero signals in an $N$-dimensional complex vector space $\mathcal{W}$ such that $\mathcal{F}$ spans $\mathcal{W}$. We refer to $\mathcal{F}$ as a *frame* or a *dictionary* for $\mathcal{W}$. For $r \in \mathcal{W}$ there are possibly infinite ways to represent r as a linear combination of the elements of $\mathcal{F}$. In this paper, we are interested in sparse representations of r with the lowest number of non-zero coefficients (referred to as the $\mathcal{L}_0$ norm of the representation vector). In fact, sparse representations have recently received wide attention because of their numerous potential applications. Such applications include Magnetic Resonance Imaging, where only a partial set of measurements are available to describe an object [1]; compression using overcomplete dictionaries; separation of images into disjoint signal types, etc. (Please see [5] and the references therein).

If the signal to be represented is known to have a sparse representation, the question of interest is to find the exact sparsest representation in terms of the dictionary elements. This problem will be referred to as the noiseless coding problem. The difficulty of this problem has caused many researchers to look for approximations to the solution. Two most commonly used methods are the orthogonal matching

M. Akçakaya and V. Tarokh are with the Division of Engineering and Applied Sciences, Harvard University, Cambridge, MA, 02138. (e-mails: {akcakaya, vahid}@deas.harvard.edu)

pursuit (OMP) and basis pursuit (BP). OMP is a greedy algorithm, which generalizes the classical orthonormal basis algorithm. The OMP algorithm starts with the residual signal at step zero set to be the original signal r. Then at each step $i$, the dictionary element that has the highest correlation with the residual signal is selected. The residual signal at step $i \geq 1$ is then updated to be the projection of the residual signal at time $i-1$ on the orthogonal complement space of the subspace spanned by the dictionary elements chosen up to and including the stage $i$ [14]. In contrast, the basis pursuit (BP) algorithm is based on a linear programming approach to the sparse representation problem, where instead of minimizing the number of nonzero coefficients in the approximation, minimization of the sum of the absolute values of the coefficients (i. e. the $\mathcal{L}_1$ norm of the representation vector) is the objective [4]. Both algorithms can be applied to arbitrary dictionaries, and little attention has been paid to the construction of dictionaries that support simple algorithms for the computation of sparse representations. However, the construction of such dictionaries and their importance may be evident from various applications [5]. This motivates our studies in this paper, where we consider two important cases of this problem namely the *noiseless* and *noisy* sparse representation problems.

The *noiseless sparse representation problem* considers the case when $r \in \mathcal{W}$ and it is known in advance that the signal r has a sparse representation in terms of the elements of the frame $\mathcal{F}$. In this case, the goal is to find the solution to the sparse representation problem in real time. This is a problem commonly encountered in signal theory. For instance, when the underlying frame is the Fourier basis, then the classical problem of finding the Fourier expansion coefficients is of immense interest. In this case, a question of fundamental importance is the fundamental limits on sparsity of r for which a unique noiseless representation exists, and the construction of the frames that achieve these fundamental bounds and support real time solutions. We will provide a solution to this problem in this paper.

The *noisy sparse representation problem* considers the case when $r \in \mathcal{W}$ is not known to have a sparse representation. In this case, the signal r cannot necessarily be represented in terms of the elements of the frame $\mathcal{F}$ in a sparse manner, and any such sparse representation suffers from some *distortion*. The objective in this case is to trade-off sparsity for distortion. Let the redundancy of $\mathcal{F}$ be $r - 1$, where $r = M/N$. We will study the trade-off between sparsity, distortion and redundancy, a problem of fundamental importance. Another important problem is to construct frames $\mathcal{F}$ for which not only these trade-offs can be achieved, but also the underlying sparse representations can be found in real time, and this is currently being investigated.

The outline of this paper is given next. In Section II, we provide a solution to the noiseless sparse representation problem. We will construct an explicit frame, which we refer to as the Vandermonde frame, for which the noiseless sparse representation problem can be found uniquely using $O(N^2)$ operations, as long as the number of non-zero coefficients of the sparsest representation of $\mathbf{r}$ over the frame is $\epsilon N$ for $0 \leq \epsilon \leq 0.5$. We will also argue that $\epsilon \leq 0.5$ cannot be relaxed, without violating uniqueness. In Section III, we consider the noisy sparse representation problem and propose a statistical approach to this problem. We compute a lower bound on the trade-off between sparsity, distortion and redundancy for any frame $\mathcal{F}$. Finally in Section IV, we will make our conclusions and provide directions for future research.

## II. THE NOISELESS SPARSE REPRESENTATION PROBLEM

As in the previous section, consider a frame $\mathcal{F} = \{\boldsymbol{\phi}_1, \boldsymbol{\phi}_2, \cdots, \boldsymbol{\phi}_M\}$ of $M \geq N$ non-zero vectors that span an $N$ dimensional subspace $\mathcal{W} \leq \mathbb{C}^M$. Any vector $\mathbf{r}$ in $\mathcal{W}$ can be written in (possibly a non-unique) way as the sum of the elements of $\mathcal{F}$. Let $\mathbf{r} = \sum_{i=1}^{M} c_i \boldsymbol{\phi}_i$ be such a representation. We define $\|\mathbf{r}\|_{0,\mathcal{F}}$ to be the smallest number of non-zero coefficients of any such expansion. Also, for an arbitrary vector $\mathbf{c} = (c_1, c_2, \cdots, c_M) \in \mathbb{C}^M$, we define $\|\mathbf{c}\|_0$ to be the number of non-zero elements of $\mathbf{c}$. Thus $\|\mathbf{r}\|_{0,\mathcal{F}}$ is simply the $\min(\|\mathbf{c}\|_0)$ over all possible expansions $\mathbf{c}$ of $\mathbf{r}$ as above. A main problem of interest is

- *The Most Compact Representation (MCR) Problem*: Given $\mathcal{F}$ a frame spanning $\mathcal{W}$, and $\mathbf{r} \in \mathcal{W}$ find an expansion $\mathbf{r} = \sum_{i=1}^{M} c_i \boldsymbol{\phi}_i$ for which $\mathbf{c} = (c_1, c_2, \cdots, c_M)$ has minimum $\|\mathbf{c}\|_0$.

Let $\mathbf{F}$ be the matrix whose rows are the elements of the frame $\mathcal{F}$. Then the MCR Problem can be restated as:

$$\min_{\mathbf{c} \in \mathbb{C}^M} \|\mathbf{c}\|_0 \quad \text{s.t.} \quad \mathbf{r} = \mathbf{c}\mathbf{F} \tag{1}$$

This optimization problem is in general difficult to solve. In this light, much attention has been paid to solutions minimizing $\|\mathbf{c}\|_1 = \sum_{i=1}^{M} |c_i|$ instead, and then establishing criteria under which the minimizing $\mathbf{c}$ also solves the MCR Problem [1], [9].

In this section, we will take a different approach. We will construct an explicit frame $\mathcal{F}$ for which the following problem:

- *Decoding Problem*: Whenever $\mathbf{r}$ has a representation $\mathbf{c}$ with $\|\mathbf{c}\|_0 = \epsilon N$, for $0 \leq \epsilon \leq 0.5$, then find $\mathbf{c}$, can be solved with a *unique* answer in running time $O(N^2)$.

### A. Connection with Error Correcting Codes

In solving the MCR problem using the above approach, we make a simple albeit fundamental connection between solutions to the MCR Problem and error correcting coding/decoding. Such connections have also been made by a number of other authors who have realized connections between frames and linear codes defined over the field of complex numbers [9]. Inspired by this connection and the theory of algebraic coding/decoding, we construct frames that generalize Reed-Solomon codes using Vandermonde matrices. Under the assumption that $\|r\|_{0,\mathcal{F}} \leq N/2$, a generalized Reed-Solomon decoding algorithm (which corrects up to half of the minimum distance bound) can find the solution to the decoding problem and the MCR problem. Such decoding algorithms and their improvements are well-known in the coding theory literature.

Consider a frame $\mathcal{F} = \{\boldsymbol{\phi}_1, \boldsymbol{\phi}_2, \cdots, \boldsymbol{\phi}_M\}$ of $M$ non-zero vectors that span an $N$ dimensional subspace $\mathcal{W} \subseteq \mathbb{C}^M$ as above. Consider:

$$\mathcal{V} = \{\mathbf{d} = (d_1, d_2, \cdots, d_M) \in \mathbb{C}^M : \sum_{i=1}^{M} d_i \boldsymbol{\phi}_i = 0\}$$

The vector space $\mathcal{V}$ is clearly an $M - N$ dimensional subspace of $\mathbb{C}^M$. If $\mathbf{r} \in \mathcal{W}$ can be represented by $\mathbf{c}$ with respect to the above frame $\mathcal{F}$, then all possible representations of $\mathbf{r}$ are given by $\mathbf{c} - \mathcal{V} = \{\mathbf{c} - \mathbf{d} \mid \mathbf{d} \in \mathcal{V}\}$. Thus the problem of finding the sparsest representation of $\mathbf{r}$ is equivalent to finding $\mathbf{d} \in \mathcal{V}$ which minimizes $\|\mathbf{c} - \mathbf{d}\|_0$. If one thinks of $\mathcal{V}$ as a linear code defined over the field of complex numbers, and of $\mathbf{r}$ as the received word, the MCR Problem is equivalent to finding the error vector $\mathbf{e} = \mathbf{c} - \mathbf{d}$ of minimum (Hamming weight) $\|\mathbf{e}\|_0$ over all the codewords $\mathbf{d} \in \mathcal{V}$. Problems of this nature have been widely studied in the language of coding theory, however these codes are typically defined over finite fields. The main contribution of this paper is the observation that complex analogues of the Reed-Solomon codes can be constructed from Vandermonde matrices and the associated sparse representations can be computed using many well-known Reed-Solomon type decoding algorithms.

### B. Vandermonde Frames

Consider the matrix given below:

$$\mathbf{A} = \begin{pmatrix} 1 & z_1 & z_1^2 & \cdots & \cdots & z_1^{N-1} \\ 1 & z_2 & z_2^2 & \cdots & \cdots & z_2^{N-1} \\ 1 & z_3 & z_3^2 & \cdots & \cdots & z_3^{N-1} \\ \vdots & \vdots & \ddots & \ddots & \ddots & \vdots \\ 1 & z_M & z_M^2 & \cdots & \cdots & z_M^{N-1} \end{pmatrix} \tag{2}$$

where $z_i, \ i = 1, 2, \cdots, M$ are *distinct, non-zero* complex numbers. The following

- *Condition I*: Any arbitrary set of $N$ distinct rows of $\mathbf{A}$ are linearly independent

holds. This is clear since any such $N$ rows form a Vandermonde matrix with non-zero determinant.

We define our frame $\mathcal{F}$ to consist of the rows of $\mathbf{A}$, i.e.

$$\mathcal{F} = \{\boldsymbol{\phi}_j = (1, z_j^1, z_j^2, \ldots, z_j^{N-1}) \text{ for } j = 1, \cdots, M\} \tag{3}$$

and refer to it as a *Vandermonde* frame.

Let $\mathcal{W}$ be the $N$ dimensional subspace spanned by the elements of $\mathcal{F}$. The subspace $\mathcal{V}$, as defined above, is given by the vectors $\mathbf{d} = (d_1, d_2, \cdots, d_M)$ for which

$$d^1 = d^2 = \cdots = d^N = 0, \tag{4}$$

where

$$d^i = \sum_{j=1}^{M} d_j z_j^{i-1}. \qquad (5)$$

Clearly the subspace $\mathcal{V}$ is $M - N$ dimensional. Furthermore if $\mathbf{w} \in \mathcal{W}$ and $\mathbf{w} = \sum_{i=1}^{M} c_i \boldsymbol{\phi}_i$, then

$$\mathbf{w} = (w^1, w^2, \cdots, w^N), \qquad (6)$$

where

$$w^i = \sum_{j=1}^{M} c_j z_j^{i-1}. \qquad (7)$$

The subspace $\mathcal{V}$ has the following interesting property.

*Lemma 2.1:* For any non-zero vector $\mathbf{v} \in \mathcal{V}$, we have $\|\mathbf{v}\|_0 > N$. Moreover, there exist vectors $\mathbf{v} \in \mathcal{V}$ with $\|\mathbf{v}\|_0 = N + 1$.

*Proof:* Suppose that $\mathbf{v} \in \mathcal{V}$ is non-zero and $\|\mathbf{v}\|_0 \leq N$. Let the nonzero elements of $\mathbf{v}$ occur in locations $\{j_1, j_2, \cdots, j_w\}$ where $w = \|\mathbf{v}\|_0 \leq N$. Then rows $\{j_1, j_2, \cdots, j_w\}$ of matrix $\mathbf{A}$ are dependent. This violates Condition I. To observe that there are vectors $\mathbf{v} \in \mathbf{V}$ with $\|\mathbf{v}\|_0 = N + 1$, we have to only exhibit a linear dependence between $N + 1$ rows of $\mathbf{A}$. This is trivial since any arbitrary $N + 1$ rows of $\mathbf{A}$ are linearly dependent. $\qquad \square$

In the language of algebraic coding theory, the subspace $\mathcal{V}$ is a maximum distance separable (MDS) linear code of length $M$, dimension $M - N$ and minimum distance $N + 1$. In fact as we will see from our decoding algorithm, this subspace provides complex analogues of the Reed-Solomon codes.

### C. Uniqueness of The Sparsest Representation

Next, let a vector $\mathbf{r} \in \mathcal{W}$ be given and we are given that $\|\mathbf{r}\|_{0,\mathcal{F}} \leq N/2$. We will first prove the following important Lemma.

*Lemma 2.2:* Given that $\|\mathbf{r}\|_{0,\mathcal{F}} \leq N/2$, the solution to the Decoding Problem is unique.

*Proof:* Let $\mathbf{r} = \sum_{j=1}^{M} c_j \boldsymbol{\phi}_j$ and $\mathbf{r} = \sum_{j=1}^{M} d_j \boldsymbol{\phi}_j$ be two solutions to the Decoding Problem with $\|\mathbf{c} = (c_1, \ldots, c_M)\|_0 \leq N/2$ and $\|\mathbf{d} = (d_1, \cdots, d_M)\|_0 \leq N/2$. Then $\mathbf{c} - \mathbf{d} \in \mathcal{V}$ and $\|\mathbf{c} - \mathbf{d}\|_0 \leq \|\mathbf{c}\|_0 + \|\mathbf{d}\|_0 \leq N$. By Lemma 2.1, $\mathbf{c} - \mathbf{d} = 0$ and $\mathbf{c} = \mathbf{d}$. $\qquad \square$

The importance of Lemma 2.2 follows from the following obvious albeit fundamental observation. If in the decoding algorithm, we were interested in finding all the representations $\mathbf{c}$ of $\mathbf{r}$ with $\|\mathbf{c}\|_0 \leq (N+1)/2$, then the solution was not necesarily unique. This non-uniqueness can be seen from Lemma 2.1. Because there exists $\mathbf{v} \in \mathcal{V}$ with $\|\mathbf{v}\|_0 = N + 1$, one can easily construct two distinct representations of the same vector in $\mathcal{W}$ both having $(N+1)/2$ non-zero coefficients (provided that $(N + 1)/2$ is an integer). Thus the bound in the Decoding Problem cannot be improved assuming that the algorithm is to output a unique solution.

We note that the uniqueness of the solution to the Decoding Problem is not necessary when considered from the point of view of frame theory. In fact, our proposed decoding framework may be readily generalized using well-known methods in coding theory to design algorithms that list all possible compact representations of a given vector $\mathbf{r} \in \mathcal{W}$. Such algorithms are known as *list decoding algorithms* in the literature [11].

### D. The Decoding Algorithm

We next provide a polynomial time algorithm that outputs the sparsest representation of $\mathbf{r}$ under the assumption that $\|\mathbf{r}\|_{0,\mathcal{F}} \leq N/2$. Let $\mathbf{r} = \sum_{j=1}^{M} c_j \boldsymbol{\phi}_j$ be an arbitrary representation of $\mathbf{r}$ in this frame. A candidate $\mathbf{c}$ can be easily computed using $O(N^2)$ operations. For example if we let $c_{N+1} = \cdots = c_M = 0$, then $c_1, c_2, \cdots, c_N$ can be computed by multiplying the inverse of a Vandermonde matrix (that can be once computed off-line) by $\mathbf{r}$, requiring at most $2N^2$ operations. We fix the representation $(c_1, c_2, \cdots, c_M)$ of $\mathbf{r}$ and seek to compute the most compact description $(e_1, e_2, \cdots, e_M)$ of $\mathbf{r} = \sum_{j=1}^{M} e_j \boldsymbol{\phi}_j$ in this frame with $\mathbf{e} = \|(e_1, e_2, \cdots, e_M)\|_0 \leq N/2$. Clearly

$$(c_1, c_2, \cdots, c_M) = \mathbf{e} + \mathbf{d}, \qquad (8)$$

where $\mathbf{d} = (d_1, d_2, \cdots, d_M) \in \mathcal{V}$. For any $i = 1, 2, \cdots$, let

$$d^i = \sum_{j=1}^{M} d_j z_j^{i-1},$$

and

$$e^i = \sum_{j=1}^{M} e_j z_j^{i-1}, \qquad (9)$$

$$c^i = \sum_{j=1}^{M} c_j z_j^{i-1}, \qquad (10)$$

then by Equation (4), we have

$$c^i = e^i \qquad (11)$$

for $i = 1, 2, \cdots, N$. Thus $e^i, i = 1, 2, \cdots, N$ can be computed using at most $2MN$ operations.

Let the nonzero elements of $\mathbf{e} = (e_1, e_2, \cdots, e_M)$ be in $i_1, i_2, \cdots, i_w$ where $w \leq N/2$. For $j = 1, 2, \cdots, w$, let $X_j = z_{i_j}$ and $Y_j = e_{i_j}$. The following Lemma gives the analogue of the Key Equation in Reed-Solomon decoding [7].

*Lemma 2.3:* Define

$$\sigma[z] = \prod_{i=1}^{w} (1 - X_i z), \qquad (12)$$

$$\omega[z] = \sum_{i=1}^{w} Y_i \prod_{j=1, j \neq i}^{w} (1 - X_j z), \qquad (13)$$

$$S[z] = \sum_{i=1}^{\infty} e^i z^{i-1}, \qquad (14)$$

then

$$\omega[z] = S[z] \sigma[z], \qquad (15)$$

anywhere in the disk $|z| < \min_{1 \leq j \leq M} (|z_j|^{-1})$.

*Proof:* Although the proof is given in [7], for completeness we repeat the proof here. Clearly

$$\frac{\omega[z]}{\sigma[z]} = \sum_{i=1}^{w} \frac{Y_i}{1 - X_i z}. \tag{16}$$

Under the assumption of $|z| < \min_{1 \le j \le M}(|z_j|^{-1})$, we have

$$\frac{1}{1 - X_i z} = \sum_{j=0}^{\infty} (zX_i)^j.$$

Replacing this in Equation (16), we have

$$\frac{\omega[z]}{\sigma[z]} = \sum_{i=1}^{w} \sum_{j=1}^{\infty} Y_i X_i^{j-1} z^{j-1}. \tag{17}$$

Clearly $e^j = \sum_{i=1}^{w} Y_i X_i^{j-1}$. Thus the result follows. $\square$

Since $\deg(\omega[z]) \le \epsilon N - 1 \le N/2 - 1$ and $\deg(\sigma[z]) = \epsilon N \le N/2$ only $e^1, e^2, \cdots, e^N$ are needed to compute $\omega[z]$ and $\sigma[z]$ from the above (for instance by solving a linear system of equations for the coefficients of $\omega[z]$ and $\sigma[z]$). It is well-known that this task can be achieved more efficiently using the Euclid division algorithm [7]. In fact, letting $S_1[z] = \sum_{j=1}^{N} e^j z^{j-1}$ one can write:

$$\omega[z] = S_1[z]\sigma[z] \mod (z^N)$$

for all $z \in \mathbb{C}$. The computation of $\omega[z]$ and $\sigma[z]$ can be performed using the Euclid division algorithm as described for instance in [7] (Section 9, Chapter 12). The number of operations required for the execution of this algorithm is clearly $O(MN)$.

Once $\sigma[z]$ and $\omega[z]$ are found, we first compute $\sigma[z]$ for $z_1^{-1}, z_2^{-1}, \cdots, z_M^{-1}$. This step only requires $O(\epsilon N^2)$ computations (since the required powers of $z_j$, $j = 1, 2, \cdots, M$ must only be once computed off-line). In this way, the roots $z_{i_1}^{-1}, \cdots, z_{i_w}^{-1}$ of $\sigma[z]$ (and hence the locations of non-zero elements of $\mathbf{e}$) can be found. The values $e_{i_1}, \cdots, e_{i_w}$ can then be found using the formula (attributed to Forney)

$$Y_j = \frac{\omega(X_j^{-1})}{\prod_{i=1, i \ne j}^{w}(1 - X_i X_j^{-1})} = \frac{X_j \omega(X_j^{-1})}{\sigma'[X_j^{-1}]}, \tag{18}$$

where $\sigma'[z]$ is the derivative of $\sigma[z]$.

In conclusion the vector $\mathbf{e}$ giving the most compact representation of $\mathbf{r}$ can be computed with complexity $O(N^2)$.

We note that the explicit construction in Section II improves constructively on the required sparsity factor of other existing techniques [3] that provide a solution to the MCR problem. However, the basis pursuit and OMP algorithms also work for the noisy sparse representation problem. It is not immediately clear how to solve the noisy sparse representation problem for the Vandermonde frames and this topic is currently being investigated.

## III. The Noisy Sparse Representation Problem

The noisy sparse representation problem considers the case when $\mathbf{r} \in \mathcal{W}$ is not known to have an exact sparse representation. Then $\mathbf{r}$ cannot necessarily be represented in terms of the elements of the frame $\mathcal{F}$ in a sparse manner, and any such sparse representation suffers from some *distortion*. In addressing this distortion, since any scaling of $\mathbf{r}$ by a factor of $\alpha \ne 0$ changes the distortion in a sparse representation of $\mathbf{r}$ in terms of a set of given vectors by a factor of $|\alpha|^2$ and does not affect the sparsity, it can be assumed without loss of generality, that $||\mathbf{r}||_2 = \sqrt{N}$.

Two classes of noisy representation problems have been considered in the literature, namely *bounded distance sparse decoding* (BDSD) and *sparse minimum distance decoding* (SMDD) [12]. These are formulated as:

$$\min_{\mathbf{c} \in \mathbb{C}^M} ||\mathbf{c}||_0 \quad \text{s.t.} \ ||\mathbf{r} - \mathbf{cF}||_2^2 \le \delta ||\mathbf{r}||^2 \quad \text{(BDSD)} \tag{19}$$

or

$$\min_{\mathbf{c} \in \mathbb{C}^M} ||\mathbf{r} - \mathbf{cF}||_2^2 \quad \text{s.t.} \ ||\mathbf{c}||_0 \le \epsilon N \quad \text{(SMDD)} \tag{20}$$

Both BDSD and SMDD problems have been studied in an $\mathcal{L}_1$ setting [2], [5] or by using OMP methods [13]. Almost all the existing research are based on a worst case criterion, where the maximum distortion over all vectors $\mathbf{r}$ is the underlying measure of performance.

The SMDD and BDSD problems are intimately related. In particular, if the SMDD problem can be solved in polynomial time, then so can the BDSD problem. In fact the threshold on the value of $||\mathbf{c}||_0$ can be decreased from $N$ to $0$ in $N$ applications (or in $\log_2(N)$ applications, with a binary search) of the polynomial time algorithm for the SMDD problem, until the minimum distortion exceeds the required threshold for the BDSD problem.

For a given $\mathbf{r}$, the SMDD problem finds the closest (in the Euclidean distance sense) sparse representation $\sum_{i \in \mathcal{I}} c_i \phi_i$ (with $|\mathcal{I}| \le \epsilon N$). In order to quantify the performance of SMDD, we propose an *average distortion approach*. Such measures are motivated by information theory and by the fact that we seek frames for which the SMDD works well for *typical* signals. In fact, if a given vector $\mathbf{r}$ is selected according to a given distribution on the hypersphere of radius $\sqrt{N}$, an appropriate frame must be designed to reduce the average distortion of SMDD. If no knowledge of $\mathbf{r}$ is at hand, it is natural to assume that $\mathbf{r}$ is distributed *uniformly* on the complex hypersphere of radius $\sqrt{N}$ centered at the origin, and this assumption will be used throughout the rest of this paper. In formal words, for any frame $\mathcal{F}$, and sparsity factor $0 \le \epsilon \le 1$, our measure of performance of $\mathcal{F}$ is given by

$$D(\mathcal{F}) = \frac{1}{N} \mathbb{E}_{\mathbf{r}} \min ||\mathbf{r} - \mathbf{cF}||^2,$$

where the minimum is taken over all representations $\mathbf{c}$ of $\mathbf{r}$ with $||\mathbf{c}||_0 \le \epsilon N$ and the expectation is for $\mathbf{r}$ uniformly distributed on the $N$ dimensional complex hypersphere of radius $\sqrt{N}$ centered at the origin.

### A. Trade-off Between Sparsity, Redundancy and Distortion

Let $M = rN$, where $r - 1$ is the redundancy, and let $L = \epsilon N$, where $\epsilon$ is the required sparsity. Let $T = \binom{M}{L}$.

Consider all the $L$-dimensional subspaces of $\mathcal{W}$ that are spanned by all subsets of size $L$ of $\{\phi_j\}_{j \in \mathcal{I}_k}$. There are $T_* \le T$ distinct $L$-dimensional such subspaces denoted by $\{\mathcal{P}_k, k = $ $1, 2, \cdots, T_*$}. Given a vector $\mathbf{r}$ on the $N$ dimensional complex hypersphere, the SMDD algorithm find the closest $\mathcal{P}_k, k = 1, 2, \cdots, T_*$ to $\mathbf{r}$. In other words, it minimizes $||\mathbf{r} - \Pi_{\mathcal{P}_k}\mathbf{r}||^2$, where $\Pi_{\mathcal{P}_k}$ is the projection operator onto $\mathcal{P}_k$.

Using this geometric interpretation, we will find a lower bound on the distortion as a function of sparsity and redundancy of any frame $\mathcal{F}$. To this end, we define an $L$-dimensional complex generalized cap of radius $\sqrt{\rho}$ around an $L$-dimensional plane $\mathcal{P}_k$ as

$$GC_L(\rho, \mathcal{P}_k) = \{\mathbf{x} \in \mathbb{S}^N : ||\mathbf{x} - \Pi_{\mathcal{P}_k}\mathbf{x}||^2 \le \rho\} \quad (21)$$

where $\mathbb{S}^N$ is the $N$ dimensional complex unit hypersphere

$$\{\mathbf{x} \in \mathbb{C}^N : ||\mathbf{x}||^2 = 1\}.$$

If we are only interested in the radius of the generalized cap, but not the specific plane, we will use the notation $GC_L(\rho)$.

In order to calculate the quantity of interest, $\frac{1}{N}\mathbb{E}_\mathbf{r}\left(\min_{1 \le k \le T_*} ||\mathbf{r} - \Pi_{\mathcal{P}_k}\mathbf{r}||^2\right)$ for $\frac{\mathbf{r}}{\sqrt{N}}$ uniformly distributed on $\mathbb{S}^N$, we need to know the distribution of $\min_k d^2(\mathbf{x}, \mathcal{P}_k)$, where $d^2(\mathbf{x}, \mathcal{P}_k) \triangleq ||\mathbf{x} - \Pi_{\mathcal{P}_k}\mathbf{x}||^2$ and $\mathbf{x}$ is uniformly distributed on $\mathbb{S}^N$. Clearly, for any given $\mathbf{x} \in \mathbb{S}^N$, we have

$$\mathbb{P}(\min_k d^2(\mathbf{x}, \mathcal{P}_k) \le \eta)$$
$$= \mathbb{P}(\text{There exists a plane within distance } \sqrt{\eta} \text{ of } \mathbf{x})$$
$$= \mathbb{P}(\mathbf{x} \text{ is in the area covered by the generalized}$$
$$\text{caps of radius } \sqrt{\eta})$$
$$= \mathbb{P}\left(\mathbf{x} \in \bigcup_{k=1}^{T_*} GC_L(\mathcal{P}_k, \eta) \mid \mathbf{x}\right)$$

Since $T_* \le T$

$$\mathbb{P}\left(\mathbf{x} \in \bigcup_{k=1}^{T_*} GC_L(\mathcal{P}_k, \eta)\right)$$
$$\le \sum_{k=1}^{T_*} \mathbb{P}\left(\mathbf{x} \in GC_L(\mathcal{P}_k, \eta)\right) \le T\frac{\mathcal{A}(GC_L(\eta))}{\mathcal{A}(\mathbb{S}^N)} \quad (22)$$

Thus

$$\mathbb{P}(\min_k d^2(\mathbf{x}, \mathcal{P}_k) \le \eta) \le T\frac{\mathcal{A}(GC_L(\eta))}{\mathcal{A}(\mathbb{S}^N)},$$

and

$$\mathbb{P}(\min_k d^2(\mathbf{x}, \mathcal{P}_k) \ge \eta) \ge \max\left(1 - T\frac{\mathcal{A}(GC_L(\eta))}{\mathcal{A}(\mathbb{S}^N)}, \; 0\right) \quad (23)$$

In order to bound

$$\mathbb{E}(\min_k d^2(\mathbf{x}, \mathcal{P}_k)) = \int_0^1 \mathbb{P}(\min_k d^2(\mathbf{x}, \mathcal{P}_k) \ge \eta) d\eta$$

We will establish a lower bound on the right hand side of Inequality (23), by estimating the value of $\eta$ for which $0 = 1 - T\mathcal{A}(GC_L(\eta))/\mathcal{A}(\mathbb{S}^N)$.

Clearly $\Pi_\mathcal{P}$ satisfies: $\Pi_\mathcal{P}^* = \Pi_\mathcal{P}$ and $\Pi_\mathcal{P}^2 = \Pi_\mathcal{P}$. It is well-known that the projection matrix $\Pi_\mathcal{P}$ can be diagonalized as $\Pi_\mathcal{P} = UDU^*$, where $U$ is a unitary matrix and $D$ is a diagonal matrix with 1's as the first $1, 2, \cdots, L$ and 0's as the rest of $N - L$ diagonal entries. Also we note that $||\mathbf{x} - \Pi_\mathcal{P}\mathbf{x}||^2 =$ $||U^*\mathbf{x} - DU^*\mathbf{x}||^2$, and $U^*\mathbf{x}$ is just a unitary transformation of $\mathbf{x}$ that has the same distribution as $\mathbf{x}$.

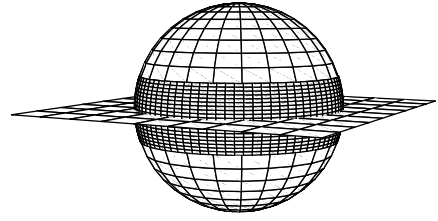

Fig. 1. A Generalized Cap in $\mathbb{R}^N$ with $N = 3$, $L = 2$

Let $\mathbf{w}$ be a vector uniformly distributed on the unit hypersphere. One way to generate $\mathbf{w}$ is to take a complex zero-mean Gaussian vector $\mathbf{z} \sim \mathcal{N}_c(0, I_N)$, where $I_N$ is the identity matrix, and let $\mathbf{w} = \frac{\mathbf{z}}{||\mathbf{z}||}$ [8] (Thm 1.5.6). Clearly $\mathbf{w}$ has the same distribution as $U^*\mathbf{w}$, and thus $||\mathbf{w} - \Pi_\mathcal{P}\mathbf{w}||^2$ has the same distribution as $||\mathbf{w} - D\mathbf{w}||^2 = |w_{L+1}|^2 + \cdots + |w_N|^2$.

It is easy to see that

$$||\mathbf{w} - D\mathbf{w}||^2 = \frac{|z_{L+1}|^2 + \cdots + |z_N|^2}{|z_1|^2 + \cdots + |z_N|^2}$$

where each $z_i \sim \mathcal{N}_c(0, 1)$ is a complex Gaussian random variable. It is well known [8] (Thm 1.5.7) that this fraction has $\beta(N - L, L)$ type distribution. Thus

$$\mathbb{P}(||\mathbf{w} - D\mathbf{w}||^2 \le \rho) = \int_0^\rho C_{N,N-L} x^{N-L-1}(1-x)^{L-1}dx$$

where $C_{N,N-L} = \frac{\Gamma(N)}{\Gamma(N-L)\Gamma(L)}$. But also

$$\mathbb{P}(||\mathbf{w} - D\mathbf{w}||^2 \le \rho) = \int_{GC_L(\rho)} \frac{1}{\mathcal{A}(\mathbb{S}^N)}d\mathbf{w} = \frac{\mathcal{A}(GC_L(\rho))}{\mathcal{A}(\mathbb{S}^N)}$$

Therefore for an $L$-dimensional generalized cap in $\mathbb{C}^N$ we have

$$\frac{\mathcal{A}(GC_L(\rho))}{\mathcal{A}(\mathbb{S}^N)} = \int_0^\rho \frac{\Gamma(N)}{\Gamma(N-L)\Gamma(L)} x^{N-L-1}(1-x)^{L-1}dx \quad (24)$$

We now prove a number of technical lemmas.

*Lemma 3.1:*

$$\int_0^\rho x^{N-L-1}(1-x)^{L-1}dx \le$$
$$\int_0^\rho \left(\frac{x}{1 + \frac{L-1}{N-L-1}x}\right)^{N-L-1}dx$$

*Proof:*

$$(1-x)^{L-1}\left(1 + \frac{L-1}{N-L-1}x\right)^{N-L-1} \le$$
$$e^{-(L-1)x}e^{(L-1)x} = 1$$

Hence we have

$$(1-x)^{L-1} \le \frac{1}{\left(1 + \frac{L-1}{N-L-1}x\right)^{N-L-1}}$$

Substituting this to the integral gives the desired result. $\quad\square$

*Lemma 3.2:* Let $f(x) = \frac{x}{1 + \frac{L-1}{N-L-1}x}$, then $f(x)$ attains its maximum for $x \in [0, \rho]$ at $\rho$.

*Proof:* By direct computation, we have

$$f'(x) = \frac{1}{(1 + \frac{L-1}{N-L-1}x)^2} > 0$$

for $x \in [0, \rho]$. Thus $f(x)$ is an increasing continuous function that attains it maximum at $\rho$. $\qquad\square$

We conclude from the above lemma that

$$\int_0^\rho \left( \frac{x}{1 + \frac{L-1}{N-L-1}x} \right)^{N-L-1} dx \leq \left( \frac{\rho}{1 + \frac{L-1}{N-L-1}\rho} \right)^{N-L-1}$$

*Lemma 3.3:*

$$\frac{\Gamma(N)}{\Gamma(N-L)\Gamma(L)} \leq (N-1)2^{(N-2)H\left(\frac{L-1}{N-2}\right)}$$

*Proof:*

$$\frac{\Gamma(N)}{\Gamma(N-L)\Gamma(L)} = \frac{(N-1)!}{(N-L-1)!(L-1)!}$$
$$= (N-1)\binom{N-2}{L-1}$$
$$\leq (N-1)2^{(N-2)H\left(\frac{L-1}{N-2}\right)}$$

$\qquad\square$

By combining these results, we obtain the following bound:

$$T\frac{\mathcal{A}(GC_L(\rho))}{\mathcal{A}(\mathbb{S}^N)} \leq T(N-1)2^{(N-2)H\left(\frac{L-1}{N-2}\right)}$$
$$\left( \frac{\rho}{1 + \frac{L-1}{N-L-1}\rho} \right)^{N-L-1} \triangleq \Lambda(\rho, N) \quad (25)$$

We define

$$\kappa_c(N) \triangleq T^{\frac{-1}{N-L-1}}(N-1)^{\frac{-1}{N-L-1}}2^{-\frac{N-2}{N-L-1}H\left(\frac{L-1}{N-2}\right)}, \quad (26)$$

then

*Lemma 3.4:* For any frame $\mathcal{F}$ of dimension $N$ and size $M$ over $\mathbb{C}$, and for a sparse representation over $\mathcal{F}$ for $\mathbf{x}$ uniformly distributed on the unit hypersphere $\mathbb{S}^N$, with at most $L = \epsilon N$ nonzero coefficients, we have

- For $L = 0$, $\mathbb{P}(\min_k d^2(\mathbf{x}, \mathcal{P}_k) \geq \rho) = 1$ for any $0 \leq \rho \leq 1$.
- For $1 \leq L \leq N-2$, the equality $\Lambda(\rho, N) = 1$ is attained at

$$\rho = \rho_0(N) \triangleq \frac{\kappa_c(N)}{1 - \frac{L-1}{N-L-1}\kappa_c(N)} \quad (27)$$

  where

$$\kappa_c(N) \triangleq T^{\frac{-1}{N-L-1}}(N-1)^{\frac{-1}{N-L-1}}2^{-\frac{N-2}{N-L-1}H\left(\frac{L-1}{N-2}\right)}$$

  and $0 \leq \rho_0(N) \leq 1$.

- For $L = N-1$, $T\mathcal{A}(GC_L(\rho))/\mathcal{A}(\mathbb{S}^N) = 1$ at

$$\rho_0(N) = 1 - \left(1 - \frac{1}{T}\right)^{\frac{1}{N-1}}.$$

- For $L = N$, $\mathbb{P}(\min_k d^2(\mathbf{x}, \mathcal{P}_k) \geq \rho) = 0$ for any $0 < \rho \leq 1$.

*Proof:* The results for $L = 0$ and $L = N$ are obvious. Thus without loss of generality, we can assume $1 \leq L \leq N-1$. Hence $T > 1$. For $1 \leq L \leq N-2$, we first claim that

$$\frac{L-1}{N-L-1}\kappa_c(N) < 1 \quad (28)$$

Clearly,

$$2^{-\frac{N-2}{N-L-1}H\left(\frac{L-1}{N-2}\right)} = \left( \frac{L-1}{N-2} \right)^{\frac{L-1}{N-L-1}} \left( \frac{N-L-1}{N-2} \right),$$

thus

$$\frac{L-1}{N-L-1}\kappa_c(N) = T^{\frac{-1}{N-L-1}}(N-1)^{\frac{-1}{N-L-1}}$$
$$\left( \frac{L-1}{N-2} \right)^{\frac{L-1}{N-L-1}} \left( \frac{L-1}{N-2} \right)$$

The claim is proved since $T^{\frac{-1}{N-L-1}} < 1$, $(N-1)^{\frac{-1}{N-L-1}} \leq 1$, and $\left( \frac{L-1}{N-2} \right)^{\frac{N-2}{N-L-1}} \leq 1$. The value of $\rho_0(N)$ and the fact that $\rho_0(N) \geq 0$ now follows from Equation (25). Since

$$T\mathcal{A}(GC_L(\rho_0(N)))/\mathcal{A}(\mathbb{S}^N) \leq \Lambda(\rho_0(N), N) = 1,$$

we have $\mathcal{A}(GC_L(\rho_0(N)))/\mathcal{A}(\mathbb{S}^N) \leq 1$ and thus $\rho_0(N) \leq 1$ as claimed.

For $L = N-1$, we directly calculate

$$\frac{\mathcal{A}(GC_L(\rho))}{\mathcal{A}(\mathbb{S}^N)} = \int_0^\rho (N-1)(1-x)^{N-2}dx = 1 - (1-\rho)^{N-1}$$

Therefore $T\mathcal{A}(GC_L(\rho))/\mathcal{A}(\mathbb{S}^N)$ becomes 1 at

$$\rho_0(N) = 1 - \left(1 - \frac{1}{T}\right)^{\frac{1}{N-1}}$$

and $0 \leq \rho_0(N) \leq 1$. $\qquad\square$

We now prove the main result of this section.

*Theorem 3.5:* For any frame $\mathcal{F}$ over $\mathbb{C}$ of dimension $N$, redundancy $r - 1 = M/N - 1$, for sparsity factor $\epsilon = L/N$, and for $\mathbf{r}$ uniformly distributed on the $N$-dimensional hypersphere of radius $\sqrt{N}$, we have

- For $L = 0$, we have $D(\mathcal{F}) = 1$.
- For $1 \leq L \leq N-2$, we have

$$D(\mathcal{F}) = \frac{1}{N}\mathbb{E}_\mathbf{r}\left( \min_{1 \leq k \leq T} ||\mathbf{r} - \Pi_{\mathcal{P}_k}\mathbf{r}||^2 \right)$$
$$\geq \rho_0(N) - \frac{\rho_0(N)^2}{\kappa_c(N)}\frac{1}{N-L}$$

  where,

$$\rho_0(N) = \frac{\kappa_c(N)}{1 - \frac{L-1}{N-L-1}\kappa_c(N)}$$

  and

$$\kappa_c(N) = T^{\frac{-1}{N-L-1}}(N-1)^{\frac{-1}{N-L-1}}2^{-\frac{N-2}{N-L-1}H\left(\frac{L-1}{N-2}\right)}.$$

- For $L = N-1$,

$$D(\mathcal{F}) = \frac{1}{N}\mathbb{E}_\mathbf{r}\left( \min_{1 \leq k \leq T} ||\mathbf{r} - \Pi_{\mathcal{P}_k}\mathbf{r}||^2 \right)$$
$$\geq \rho_0(N) - T\left( \rho_0(N) - \frac{1}{N} + \frac{1}{N}(1 - \rho_0(N))^N \right)$$

where

$$\rho_0(N) = 1 - \left(1 - \frac{1}{T}\right)^{\frac{1}{N-1}}.$$

- For $L = N$, we have $D(\mathcal{F}) = 0$.

*Proof:* The results for $L = 0$ and $L = N$ are obvious. Let $\mathbf{x} = \mathbf{r}/\sqrt{N}$, then

$$D(\mathcal{F}) = \mathbb{E}(\min_k d^2(\mathbf{x}, \mathcal{P}_k)) = \int_0^1 \mathbb{P}(\min_k d^2(\mathbf{x}, \mathcal{P}_k) \geq \eta) d\eta$$

and we have previously proved that

$$\mathbb{P}(\min_k d^2(\mathbf{x}, \mathcal{P}_k) \geq \eta) \geq \max\left(1 - T \frac{\mathcal{A}(GC_L(\eta))}{\mathcal{A}(\mathbb{S}^N)}, \; 0\right).$$

By applying Inequality (25), we have

$$\mathbb{P}(\min_k d^2(\mathbf{x}, \mathcal{P}_k) \geq \eta) \geq \max(1 - \Lambda(\eta, N), 0)$$

Combining the above and by applying Lemma (3.4), we have

$$D(\mathcal{F}) \geq \int_0^{\rho_0(N)} (1 - \Lambda(\eta, N)) d\eta$$

for $1 \leq L \leq N - 2$ and

$$D(\mathcal{F}) \geq \int_0^{\rho_0(N)} \left(1 - T \frac{\mathcal{A}(GC_L(\eta))}{\mathcal{A}(\mathbb{S}^N)}\right) d\eta$$

for $L = N - 1$. Thus for $1 \leq L \leq N - 2$

$$\begin{aligned}
D(\mathcal{F}) &\geq \int_0^{\rho_0(N)} \left[1 - T(N-1) 2^{(N-2)H\left(\frac{L-1}{N-2}\right)}\right. \\
&\qquad \left. \left(\frac{\eta}{1 + \frac{L-1}{N-L-1}\eta}\right)^{N-L-1}\right] d\eta \\
&= \rho_0(N) - \int_0^{\rho_0(N)} T(N-1) 2^{(N-2)H\left(\frac{L-1}{N-2}\right)} \\
&\qquad \left(\frac{\eta}{1 + \frac{L-1}{N-L-1}\eta}\right)^{N-L-1} d\eta \qquad (29)
\end{aligned}$$

We will next bound the last integral. Let

$$\begin{aligned}
y &= T^{\frac{1}{N-L-1}}(N-1)^{\frac{1}{N-L-1}} 2^{\frac{N-2}{N-L-1}H\left(\frac{L-1}{N-2}\right)} \frac{\eta}{1 + \frac{L-1}{N-L-1}\eta} \\
&= \frac{1}{\kappa_c(N)} \frac{\eta}{1 + \frac{L-1}{N-L-1}\eta} \qquad (30)
\end{aligned}$$

where $\kappa_c(N)$ is defined as in Equation (26). Clearly $y$ ranges from 0 to 1 in this case. Also

$$\eta = \frac{\kappa_c(N)y}{1 - y\kappa_c(N)\frac{L-1}{N-L-1}} \qquad (31)$$

Since by Equation (28), $\kappa_c(N)\frac{L-1}{N-L-1} < 1$, and $0 \leq y \leq 1$, the denominator of the above is not equal to 0 within the region of integration. Now we note that

$$d\eta = \frac{\kappa_c(N)}{\left(1 - y\kappa_c(N)\frac{L-1}{N-L-1}\right)^2} dy \triangleq g(y)dy. \qquad (32)$$

Also

$$\begin{aligned}
g(y) &= \frac{\kappa_c(N)}{\left(1 - \kappa_c(N)\frac{L-1}{N-L-1}y\right)^2} \\
&\leq \frac{\kappa_c(N)}{\left(1 - \kappa_c(N)\frac{L-1}{N-L-1}\right)^2} = \frac{\rho_0(N)^2}{\kappa_c(N)} \qquad (33)
\end{aligned}$$

since $g'(y) > 0$ for $y \in [0, 1]$. Thus the last integral in Equation (29) is bounded above by

$$\int_0^1 \frac{\rho_0(N)^2}{\kappa_c(N)} y^{N-L-1} dy = \frac{\rho_0(N)^2}{\kappa_c(N)} \frac{1}{N-L} \qquad (34)$$

which gives us

$$D(\mathcal{F}) \geq \rho_0(N) - \frac{\rho_0(N)^2}{\kappa_c(N)} \frac{1}{N-L} \qquad (35)$$

and the result for $1 \leq L \leq N - 2$.

For $L = N - 1$,

$$\begin{aligned}
D(\mathcal{F}) &\geq \int_0^1 \max\left(1 - T \frac{\mathcal{A}(GC_L(\eta))}{\mathcal{A}(\mathbb{S}^N)}, \; 0\right) d\eta \\
&= \int_0^{\rho_0(N)} 1 - T\left[1 - (1-\eta)^{N-1}\right] d\eta
\end{aligned}$$

and the result follows easily by direct integration. $\qquad \square$

We note that all the bounds attain values in $[0, 1]$, since we have integrated a function that takes values in $[0, 1]$ over a region $[0, \rho_0(N)] \subseteq [0, 1]$. Theorem 3.5 gives a fundamental limit on average distortion that *any* frame over $\mathbb{C}$ has to satisfy. We now fix $r$ and $\epsilon$ and let $N \to \infty$. The following asymptotic result follows:

*Corollary 3.6:* For any frame $\mathcal{F}$ over $\mathbb{C}$ of dimension $N$, redundancy $r - 1 = M/N - 1$, for sparsity factor $\epsilon = L/N$ and for $\mathbf{r}$ uniformly distributed on the $N$-dimensional hypersphere of radius $\sqrt{N}$, as $N \to \infty$, we have

$$D(\mathcal{F}) \geq \frac{\kappa_0(1 - \epsilon)}{1 - \epsilon \kappa_0} \qquad (36)$$

where

$$\kappa_0 = 2^{-\frac{r}{1-\epsilon}H\left(\frac{\epsilon}{r}\right)} \epsilon^{\frac{\epsilon}{1-\epsilon}} \qquad (37)$$

*Proof:* The result follows by replacing $M = rN$ and $L = \epsilon N$ in the statement of Theorem 3.5 and using

$$\frac{1}{M+1} 2^{rNH\left(\frac{\epsilon}{r}\right)} \leq T \leq 2^{rNH\left(\frac{\epsilon}{r}\right)}$$

and

$$\lim_{N \to \infty} \kappa_c(N) = \kappa_0(1 - \epsilon).$$

$\qquad \square$

It is noteworthy that the above asymptotic corollary can be proven by combining a result of Sakrison [10] with the proof method of [6], although the setting and the topic of our paper is very different from these papers.

## IV. Conclusion

In this paper, we considered approximations of signals by the elements of a frame in a complex vector space of dimension $N$. We formulated both the noiseless and the noisy sparse representation problems. For the noiseless representation problem, it is known in advance that the signal $\mathbf{r}$ has a sparse representation in terms of the elements of the frame $\mathcal{F}$. In this case, the goal is to find a solution to the sparse representation problem in real time. We provided a solution to this problem, by explicitly constructing a frame, which we referred to as the Vandermonde frame, for which the noiseless sparse representation problem can be solved uniquely using $O(N^2)$ operations, as long as the number of non-zero coefficients in the sparse representation of $\mathbf{r}$ is $\epsilon N$ for some $0 \leq \epsilon \leq 0.5$. This result improves on a result by Candes and Tao [3]. We also showed that $\epsilon \leq 0.5$ cannot be relaxed without violating uniqueness.

For the noisy sparse representation problem, we considered the case when the signal $\mathbf{r}$ cannot be represented in terms of the elements of the frame $\mathcal{F}$ in a sparse manner and noted that any such representation suffers from distortion. In this case, we established a lower bound on the trade-off between sparsity, distortion and redundancy. Our future research will focus on constructing frames for which not only these trade-offs can be achieved, but also the underlying sparse representations can be found in real time.

### Acknowledgment

The authors would like to thank Robert Calderbank, Ingrid Daubechies and David L. Donoho for many insightful discussions.